\documentclass[twocolumn,showpacs,showkeys,preprintnumbers]{revtex4}
\usepackage{amsmath}
\usepackage{amssymb}
\usepackage{graphics}

\begin{document}

\title{Evaluation of the particle numbers via the two root mean square radii
in a 2-species Bose-Einstein condensate}
\author{Y.Z.He$^{1}$, Y.M.Liu$^{2,3}$£¬ and C.G.Bao$^{1,}$}\thanks{
Corresponding author: stsbcg@mail.sysu.edu.cn, Tel:020-84037356,
Guangzhou, Xingangxi Road, 135, Zhongshan Daxue, 612-601 (code:510275)}
\affiliation{$^{1}$State Key Laboratory of Optoelectronic Materials and Technologies,
School of Physics, Sun Yat-Sen University, Guangzhou, 510275, P. R. China}
\affiliation{$^{2}$Department of Physics, Shaoguan University, Shaoguan, 512005, P. R.
China}
\affiliation{$^{3}$State Key Laboratory of Theoretical Physics, Institute of Theoretical
Physics, Chinese Academy of Sciences, Beijing, 100190, China}



\begin{abstract}
The coupled Gross-Pitaevskii equations for two-species BEC have been solved
analytically under the Thomas-Fermi approximation (TFA). Based on the
analytical solution, two formulae are derived to relate the particle numbers
$N_A$ and $N_B$  with the root mean square radii of the two kinds of atoms.
Only the case that both kinds of atoms have nonzero distribution at the center
of an isotropic trap is considered. In this case the TFA has been found to
work nicely. Thus, the two formulae are applicable and are useful for the
evaluation of $N_A$ and $N_B$.
\end{abstract}

\maketitle

Since the pioneer theoretical study by Ho and Shenoy\cite{ho96} in 1996, the
interest in two-species Bose-Einstein condensate (2-BEC) is increasing in
recent years. There are many theoretical studies
\cite{ho96,esry97,pu98,timm98,ao98,chui99,tripp00,ribo02,chui03,luo07,xu09,shi11,
gau10,nott15, scha15,inde15,roy15,luo08,polo15}. Experimentally, this system
was first achieved by Myatt, et al.\cite{myat97} in 1997. Making use of a
magnetic trap, an optical trap, or a combined magneto-optical trap, various
types of 2-BEC can be created \cite{ande05,pilc09,nemi09,wack15} (also refer
to the references listed in \cite{wack15}). In related experiments most
parameters can be known quite accurately (say, the strengths of interaction
can be precisely determined via the photo-association spectroscopy), but the
particle numbers $N_A$ and $N_B$ can not. With this background we propose an
approach which can be used for the evaluation of the particle numbers. In
details, the followings are performed.

(i) For the condensate with the A- and B-atoms, we have derived two formulae
to relate the two root mean square radii $r_{RMS}^u$ and $r_{RMS}^v$, respectively,
to the parameters involved in the experiments. Since the root mean square radii
are observable, these two formulae are useful for the determination or refinement
of the parameters.

(ii) We have find out the border separating the whole parameter-space into two
subspace for miscible and immiscible phases, respectively. The determination of
the border provides a base for plotting the phase-diagrams,\cite{llyb} and
therefore helps to understand intuitively the inherent physics.

(iii) Since we have introduced the Thomas-Fermi approximation (TFA) in the
derivation (in which the kinetic energy has been neglected), we have performed a
numerical calculation to evaluate the error caused by the TFA. In this way the
applicability of the two formulae is clarified.

Let the masses of the A- and B-atoms be $m_A$ and $m_B$. These cold
atoms are subjected to the isotropic parabolic potentials
$\frac{1}{2}m_S\omega_S^2 r^2$ ($S=A$ or $B$). We introduce a mass $m$ and a
frequency $\omega$. $\hbar\omega$ and $\lambda\equiv\sqrt{\hbar/(m\omega)}$
are used as units for energy and length in this paper. Then, the intra-species
interaction $V_S=c_S\sum_{i<i'}\delta(\mathbf{r}_i-\mathbf{r}_{i'})$, and the
inter-species interaction $V_{AB}=c_{AB}\sum_{i<j}\delta (\mathbf{r}_i-\mathbf{r}_j)$.
Their spin-degrees of freedom are considered as being frozen. The ground state
(g.s.) $\Psi_{\mathrm{gs}}$ is assumed to have the following form
\begin{equation}
 \Psi_{\mathrm{gs}}
  =  \prod_{j=1}^{N_A}
     \frac{u(r_j)}{\sqrt{4\pi }r_j} \
     \prod_{k=1}^{N_B}
     \frac{v(r_k)}{\sqrt{4\pi}r_k}
 \label{eq1}
\end{equation}
where $u(r)$ and $v(r)$ are for the A- and B-atoms, respectively, and they
are most advantageous to binding. We further introduce
$\gamma_S\equiv\frac{m_s}{m}(\frac{\omega s}{\omega})^2$ and a set of four
parameters $\alpha _1\equiv N_A|c_A|/(4\pi \gamma _A)$, $\beta _1\equiv
N_B|c_{AB}|/(4\pi \gamma _A)$, $\alpha _{2}\equiv N_B|c_B|/(4\pi
\gamma _B)$, and $\beta _{2}\equiv N_A|c_{AB}|/(4\pi \gamma _B)$,
where $|c_A|$ is dimensionless and is the value of $c_A$\ in the new units, etc..
This set is called the weighted strengths (W-strengths). Under the TFA, the coupled Gross-Pitaevskii equations (CGP) for
$u$ and $v$ in the dimensionless form appear as
\begin{eqnarray}
 &&( \frac{r^2}{2}
    +\alpha_1\frac{u^2}{r^2}
    +\beta_1\frac{v^2}{r^2}
    -\varepsilon_1)u
   =0  \label{eq2} \\
 &&( \frac{r^2}{2}
    +\beta_2\frac{u^2}{r^2}
    +\alpha_2\frac{v^2}{r^2}
    -\varepsilon_2)v
   =0  \label{eq3}
\end{eqnarray}
where $\alpha_1$ and $\alpha_2$ are for the intra-species interaction and they
are considered as positive. The chemical potential for the A-atoms (B-atoms)
is equal to $\gamma_A\varepsilon_1$ ($\gamma_B\varepsilon_2$). The normalization
$\int u^2\mathrm{d}r=1$ and $\int v^2\mathrm{d}r=1$ are required. $u\geq 0$ and
$v\geq 0$ are safely assumed.

It turns out that the solutions of eqs.(\ref{eq2},\ref{eq3}) can be divided into
two phases. When both kinds of atoms have nonzero distribution at the center,
i.e. $u/r|_{r=0}>0$ and $v/r|_{r=0}>0$ (obviously, it is required that, when $r$ tends to zero, both $u$ and $v$ should tend to zero as fast as $r$), and are distributed compactly
(i.e., not distributed in disconnected regions), then it is in miscible phase.
Otherwise, in immiscible phase. For the miscible states, under the TFA, the
analytical expression for $u/r$ and $v/r$ have been given previously \cite{polo15}
but in a rather complicated form. In this paper, by introducing the W-strengths
defined ahead eq.(\ref{eq2}), we obtain a much simpler expression as given in the Appendix. Where
the kind of atoms having a narrower distribution is named as the A-atom and
described by $u$, while the other kind by $v$. The border in the parameter-space
that separates the two phases is also given in the Appendix.

With this very simple analytical expression of $u/r$ and $v/r$, it is straight
forward to obtain the root mean square radii from the definitions
$r_{\mathrm{RMS}}^u\equiv(\int u^2 r^2 \mathrm{d}r)^{1/2}$ and
$r_{\mathrm{RMS}}^v\equiv(\int v^2 r^2 \mathrm{d}r)^{1/2}$. Thus we have
\begin{eqnarray}
 &&(r_{\mathrm{RMS}}^u)^2
   =  \frac{3}{7}
      [ \frac{15(\alpha_1\alpha_2-\beta_1\beta _2)}{\alpha _2-\beta _1}]^{2/5}
 \label{eq4} \\
 && \beta_2(r_{\mathrm{RMS}}^u)^2
   +\alpha_2(r_{\mathrm{RMS}}^v)^2
   = 15^{2/5}
     \frac{3}{7}
     (\alpha_2+\beta_2)^{7/5}
 \label{5}
\end{eqnarray}
Obviously, when all the parameters are known except $N_A$ and $N_B$, and
when $r_{\mathrm{RMS}}^u$ and $r_{\mathrm{RMS}}^v$ have been measured, $N_A$ and
$N_B$ can be known from eqs.(\ref{eq4},\ref{5}).

\begin{figure}[tbp]
\centering \resizebox{0.95\columnwidth}{!}{\includegraphics{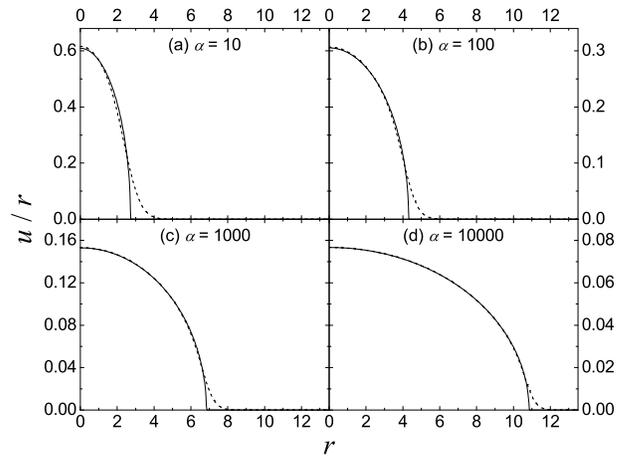} }
\caption{(color on line) The wave function $u/r$ under TFA (solid line) and
obtained from exact calculation (dash line).}
\label{fig:1}
\end{figure}

Since the TFA has been adopted, we have to evaluate the deviation caused by
the TFA. For this aim, we go to the one-species BEC. When $\hbar\omega_S$
and $\lambda_S\equiv\sqrt{\hbar/(m_S\omega_S)}$ are used as units, the
dimensionless Gross-Pitaevskii equation is
\begin{equation}
 (-\frac{d^2}{2\mathrm{d}r^2}+\frac{1}{2}r^2+\alpha \frac{u^2}{r^2})u
 =  \varepsilon u
 \label{eq8}
\end{equation}
where $\alpha=N_S|c_S|/(4\pi)$. Under the TFA,
$u/r=\sqrt{\frac{15}{2r_0^3}}(1-\frac{r^2}{r_0^2})^{1/2}$, where
$r_0=(15\alpha)^{1/5}$. The root mean square radius
$R_{\mathrm{TFA}}=\sqrt{3/7}(15\alpha)^{1/5}$. Let the radius obtained from the
exact solution of eq.(\ref{eq8}) be denoted as $R_{\mathrm{exac}}$, and we define
$\alpha'=\frac{1}{15(3/7)^{5/2}}R_{\mathrm{exac}}^5$, where $\alpha'/\alpha$
measures the deviation in $\alpha$ caused by TFA. For $^{87}$Rb, the dimensionless
strength $|c_S|=0.00249\sqrt{\omega\cdot\sec}$. When $\omega=1000/sec$ as an example,
$\alpha=0.0063N_S$, where $N_S$ is assumed to be very large. For a general evaluation,
$\alpha=10$, $100$, $1000$, and $10000$ are adopted. The wave function $u/r$
obtained under TFA and from exact calculation are plotted in Fig.\ref{fig:1},
$R_{\mathrm{TFA}}$, $R_{\mathrm{exac}}$, and $\alpha'/\alpha$ are listed in Table
\ref{tab:1}.

\begin{table}[htb]
\caption{The root mean square radii and $\alpha '/\alpha $.}
\label{tab:1}
\begin{center}
\begin{tabular}{ccc|cccccc}
\hline\hline
& $\alpha$  &  & $R_{\mathrm{TFA}}$ &  & $R_{\mathrm{exac}}$    &  & $\alpha'/\alpha$   &  \\
\hline
& $10^1$    &  & $1.783$            &  & $1.883$                &  & $1.311$            &  \\
& $10^2$    &  & $2.826$            &  & $2.859$                &  & $1.059$            &  \\
& $10^3$    &  & $4.480$            &  & $4.490$                &  & $1.011$            &  \\
& $10^4$    &  & $7.100$            &  & $7.103$                &  & $1.002$            &  \\
\hline\hline
\end{tabular}
\end{center}
\end{table}

The above results demonstrate that, when the wave functions obtained under
TFA and from exact calculation overlap nicely (say, when $\alpha \geq 1000$),
$R_{\mathrm{TFA}}$ is close to $R_{\mathrm{exac}}$ and $\alpha '$ is close to
$\alpha$. It turns out that, for 2-BEC and for the case that both $u/r$ and $v/r$
are nonzero at $r=0$, the overlap of the wave functions from TFA and beyond TFA
overlap nicely (refer to Fig.1a and 1b of \cite{polo15}). Therefore, $N_A$ and
$N_B$ obtained via eqs.(\ref{eq4},\ref{5}) is reliable.

In conclusion, we have proposed an approach helpful to the determination of the
particle numbers, at least in the qualitative aspect. This approach is limited
to the case that the numbers of both kinds of atoms are huge and they have
nonzero distribution at the center. Incidentally, if the parameters other than
$N_A$ and $N_B$ are tuned to ensure $r_{\mathrm{RMS}}^u=r_{\mathrm{RMS}}^v$,
then eqs.(\ref{eq4},\ref{5}) together will lead to the equation given as eq.(6)
in the preprint \cite{llyb} which can be used to determine the ratio of the two
particle numbers.

\begin{acknowledgments}
Supported by the National Natural Science Foundation of China under Grants
No.11372122, 11274393, 11574404, and 11275279; the Open Project Program of
State Key Laboratory of Theoretical Physics, Institute of Theoretical
Physics, Chinese Academy of Sciences, China; and the National Basic Research
Program of China (2013CB933601); and Guangdong Natural Science Foundation
(2016A030313313).
\end{acknowledgments}

\section*{Appendix: Analytical solutions of the CGP under TFA for the case
related to this paper}

Let $Y_1=\frac{1}{2}(\alpha_2-\beta_1)/(\alpha_1\alpha_2-\beta_1\beta_2)$ and
$Y_2=\frac{1}{2}(\alpha_1-\beta_2)/(\alpha_1\alpha_2-\beta_1\beta_2)$. For the
case that both $u/r$ and $v/r$\ are nonzero at $r=0$ and $u/r$ has a narrower
distribution, $u$ is distributed in the domain
$(0\leq r\leq(\frac{15}{2Y_1})^{1/5}\equiv r_a)$ and appears as
\begin{equation}
 u^2/r^2 =X_1-Y_1r^2
 \label{9}
\end{equation}
where $X_1=(15/2)^{2/5}Y_1^{3/5}$. $v$ is distributed in the domain
$(0\leq r\leq\sqrt{2\varepsilon_2})$, where
$\varepsilon _2=\frac{1}{2}[15(\alpha_2+\beta_2)]^{2/5}$.

When $r\leq r_a$,
\begin{equation}
 v^2/r^2=X_2-Y_2r^2
 \label{eq10}
\end{equation}
where $X_2=(\varepsilon_2-\beta _2X_1)/\alpha_2$.

When $r_a<r\leq \sqrt{2\varepsilon_2}$
\begin{eqnarray}
 u^2/r^2 &=&0 \\
 v^2/r^2 &=&\frac{1}{\alpha_2}(\varepsilon_2-r^2/2)
 \label{eq11}
\end{eqnarray}
when $r>\sqrt{2\varepsilon_2}$, both $u$ and $v$ are zero. Thus $r_a$ and
$\sqrt{2\varepsilon_2}$ mark the borders for the A-atoms and B-atoms,
respectively. Incidentally, when $X_1$ and $X_2$ are known, $\varepsilon_1$
is related to them as $\varepsilon_1=\alpha_1 X_1+\beta_1 X_2$.

One can check directly that the above $u$ and $v$ satisfy the CGP, they are
normalized, and they are continuous at the borders (however their
derivatives are not).

Obviously, the above solution would be physically meaningful only if the
W-strengths are so preset that $Y_1>0$ and $\alpha_2+\beta_2>0$ are ensured.
To ensure $\sqrt{2\varepsilon_2}>r_a$, $Y_1\geq Y_2$ is required. Besides,
to ensure both $u/r$ and $v/r$ being $\geq 0$ at $r=0$, $X_2\geq 0$
(equivalently, $[2(\alpha_2+\beta_2)Y_1]^{2/5}\geq 2\beta _2Y_1$) is required.
These requirements imply that the suitable W-strengths will be constricted
in a subspace of the whole parameter-space.

\end{document}